\documentclass[onecolumn,aps,preprint,showpacs,amsmath,amssymb]{revtex4}
\usepackage{graphicx}
\usepackage{latexsym}
\usepackage{dcolumn} 
 \usepackage{amsmath,empheq}
\usepackage{bm}

\begin{document}
\newcommand{\eq}{\begin{equation}}                                                                         
\newcommand{\eqe}{\end{equation}}             
 
\title{Analytic solutions of a two-fluid hydrodynamic model}

\author{Imre F. Barna$^{1}$ and L. M\'aty\'as$^{2}$}
\address{ $^1$ Wigner Research Center for Physics, 
\\ Konkoly-Thege Mikl\'os \'ut 29 - 33, 1121 Budapest, Hungary \\
$^2$  Department of Bioengineering, Faculty of Economics, Socio-Human Sciences
and Engineering, Sapientia Hungarian University of Transylvania
Libert\u{a}tii sq. 1, 530104 Miercurea Ciuc, Romania} 
\date{\today}

\date{\today}

\begin{abstract} 
We investigate a one dimensional flow described with the non-compressible coupled Euler and  non-compressible 
Navier-Stokes equations in Cartesian coordinate systems. We couple the two fluids through the continuity equation where 
different void fractions can be considered. The well-known self-similar Ansatz was applied and analytic solutions were derived for both velocity and pressure field as well. 
\end{abstract}

\maketitle

\section{Introduction}
   There is no need to prove the evidence that investigation of hydrodynamic systems  
have crucial importance for human society and civilization. The second evidence is that, there are almost 
infinite variety of flows in nature or in engineering, some of them are viscous and some of them can be considered as ideal fluids. 
We may narrow the flow frames to systems of multi-phase flows where a liquid flows together with its vapor or with other 
non-condensible gas (a good example is water-air flow). 
Multi-phase flows are relevant in numerous fields like nuclear industry, hydrology chemistry or petrochemistry.   
We may refine multi-phase flows -- where large temperature gradient is present between the two phases of the media (eg. water and steam) 	
making condensation and boiling process possible -- calling it thermal-hydraulics. The most intensively studied material is the water-steam system which 
is relevant for nuclear industry.  In the last fifty years it's literature become enormous, without completeness we just mention some relevant monographs \cite{hibiki,kolev,clayton}. 
The readers who are more concentrated to the mathematics of the models should read \cite{stewart,menikoff} all models are based on gas dynamics.  
A decade ago we investigated the steam condensation induced water hammer (CIWH), which is the most complex two-phase flow including explosion-like 
condensation of hot steam to warm water. We could theoretically explain our experimentally measured 130 bar over pressure peaks which have 
2 ms pulse width \cite{imre0} for the first time. The original model was developed by Tiselj and Petelin \cite{tiselj}.  Ultimately, we investigated a proton beam induced two-phase 
flow pressure waves in mercury \cite{imre_hig}.  
Numerous thermal hydraulic models exist for two-phase flows which contain from two up to seven coupled partial differential equations (PDE) for mass, momenta and energy conservation. Unfortunately all such models are for one spatial dimension only. In the physics of multi-phase flows one of the most relevant dynamical variable is the void fraction (usually noted with $ \alpha $ ) which describes the volume ratios  (usually steam to water) of the two fluids in a given space point at a given time.  In our next model we also use the void fraction but just as a free coupling parameter between the two velocities of the two fluids.     

The second relevant field of fluid dynamics is two-fluid flows. We may define these systems where two liquids flow together with different physical parameters like, density, viscosity 
and thermal properties.  The question of superfluidity can be handled with such two-fluid models. At very low temperatures some special viscous fluids become superfluids which means, 
that they loose their internal viscosity. All technical and historical details can be found in numerous textbooks \cite{schmidt,khala,pit}. 
To put this phenomena into a wider frame, superfluidity is coupled to superconductivity and Bose-Einstein condensates in a non-trivial way. These phenomena  are sometimes called quantum fluids which (of course) can be handled with quantum hydrodynamic means \cite{wyatt,tsubota}. It is well, known from the advent of quantum mechanics, that the linear and non-linear Schr\"odinger equations can be reformulated to a hydrodynamic picture which is called the Madelung equation \cite{madel}. 
In our former studies we investigated the Madelung-Euler equations with the self-similar Ansatz and found analytic solutions for the fluid density. \cite{imre_mad1,imre_mad2}  

The original idea how to solve the superfluidity problem, namely to couple viscous and inviscid fluids came from Landau in 1941 \cite{landau}. 
Interesting aspects on the evolution of the two-fluid model related to superfluidity one can find in \cite{Ba2017}, where both the Tisza \cite{Ti38,Ti47} 
and the Landau model \cite{landau} is mentioned. Later the idea of superfluidity 
became quite widely spread  e.g. in nuclear \cite{mikdal}, high energy or astrophysics to explain exotic phases of matter \cite{hel}. 
 
It is evident from physical considerations that there are numerous ways to couple the ideal and viscous fluids together, this is done usually by their densities. 
For compressible fluids the equation of state (EOS) could also couple the dynamical equations of the two fluids via their common pressure. 
Worth to mention, that in our simple presented model we just consider an incompressible continuity equation where the velocities of the two phases are weighted with their void fraction.
This is the most simple model to couple an Euler to a Navies-Stokes equation, of course in the future we want to develop our description to more and more realistic description.  

In the following study we investigate the self-similar Ansatz \cite{sedov,zeldovich} applied the two-fluid model which describes physically relevant 
disperse or dissipate solutions. This study is organically linked to our long-term program in which we systematically goes over fundamental hydrodynamic 
systems and analyze physically relevant self-similar and traveling wave solutions. 
Till now we published about half a dozen papers  \cite{imre1,imre2,imre3,imre4} and a book chapter \cite{imre_book} in this field. 
Due to our knowledge there is no self-similar solutions known and analyzed for time-dependent two-fluid models.

\section{Theory and Results}
 Let's start with the following PDE flow system of 
\begin{eqnarray}
a\frac{\partial v_1}{\partial x} + (1-a)  \frac{\partial v_2}{\partial x} &=& 0, \nonumber \\ 
\frac{\partial v_1}{\partial t} + v_1 \frac{\partial v_1}{\partial x}  &=&
 - \frac{1}{\rho_1}\frac{\partial p}{\partial x},  \nonumber \\ 
\frac{\partial v_2}{\partial t} + v_2 \frac{\partial v_2}{\partial x}  &=& 
- \frac{1}{\rho_2}\frac{\partial p}{\partial x} + \nu\frac{\partial^2 v_2 }{\partial x^2}, 
\label{pde} 
\end{eqnarray}
where the dynamical variables are the two fluid velocities $v_1(x,t), v_2(x,t)$ and the common pressure $p(x,t)$ 
there are four additional physical constants $a,\rho_1,\rho_2$ and $\nu$ which are the void fraction, the two fluid densities and the viscosity of the second fluid. 
   
We apply the following self-similar Ansatz for the variables:  
\begin{eqnarray}
v_1(x,t) = t^{-\alpha} f(\eta), \hspace*{3mm} 
v_2(x,t) = t^{-\gamma} g(\eta),   \hspace*{3mm} 
p(x,t) = t^{-\delta} h(\eta)  
\label{ansatz}
\end{eqnarray}
with the new variable $\eta = \frac{x}{t^{\beta}}$. 
All the exponents $\alpha,\beta,\gamma,\delta $ are real numbers. (Solutions with integer 
exponents are called self-similar solutions of the first kind, non-integer exponents generate self-similar solutions of the second kind.) 
The shape functions $f,g,h$ could be any continuous functions with existing first and second continuous derivatives  and will be evaluated later on.  
The logic, the physical and geometrical interpretation of the Ansatz were exhaustively analyzed in all our 
former publications \cite{imre1,imre2,imre3,imre4, imre_book} therefore we neglect it.  

To have consistent coupled ordinary differential equation system (ODEs) for the shape functions the exponents 
have to have the following values of 
\eq 
\alpha = \beta = \gamma =  1/2, \hspace*{3mm} \delta = 1.
\eqe
Such fixed exponents were found for the multi-dimensional  incompressible  Navier-Stokes equations \cite{imre1} as well. This means that the 
pressure field has a stronger decay than the velocity fields. 

The following unequivocal ordinary differential equation (ODE) system can be obtained  
\begin{subequations}\label{eq:system}
\begin{empheq} {align}
      af' + (1-a)g'  &=  0, \label{eq:a1}
  \\
   -\frac{1}{2}f - \frac{1}{2}\eta f' + ff'  &= -\frac{h'}{\rho_1},  \label{eq:a2}
  \\
  -\frac{1}{2}g - \frac{1}{2}\eta g'  + gg' &= -\frac{h'}{\rho_2} + \nu g''.   \label{eq:a3}
\end{empheq}
\end{subequations}

Note, that unlike the tested systems so far, all three equations are total derivatives and can be integrated 
once. (All three equations are conservation equations, so this statement is straightforward, however 
our decade-long experience tells that usually only the first - the continuity -  equation has such property.)  
After some ordinary algebraic steps - which means substituting one equation into the other, and sorting the terms - 
we get the final ODE for the velocity shape function of the form of 
\eq
-\nu \rho_2 g' - \frac{\eta g \rho_2}{2} + \frac{g^2 \rho_2}{2} - c_3\rho_2 =  
-\frac{\eta \rho_1}{2}\left[ \frac{(a-1)g + c_1}{a}  \right]  +   
\frac{\rho_1}{2} \left( \frac{[ \{a-1\} g +c_1 ]}{a} \right) - c_2 \rho_2,  
\label{f_equ}
\eqe
where $c_1, c_2$ and $c_3$ are the three  integration constants of (\ref{eq:a1},\ref{eq:a2},\ref{eq:a3}).
 
All our investigated systems have a kind of hierarchy, due to the non-linearity of  the variables and the asymmetric form, 
 there is always a prior quantity which would be evaluated first.
As we see, now the viscous velocity field $g(\eta)$ comes first.  
(There is a general impenetrable many page long complex analytic solution available for (\ref{f_equ}) for the most general $ c_1, c_2, c_3 \ne 0$ case 
consisting large number of Kummer M and Kummer U functions according to the symbolic Computation Software Maple 12, which we skip now.) 
All these three integration constants mean just general  shifts in the solution function.  
For $c_1 = c_2 = c_3 = 0$ case the shape function of the velocity field is composite enough 
\eq
g  = \frac{2\nu \sqrt{ -\frac{\rho_1a - \rho_1 - \rho_2 a}{\nu \rho_2 a}  }  
 e^{ \frac{1}{4}      \frac{\eta(-\eta\rho_2 a +\eta \rho_1 a - \eta \rho_1 -2\rho_1 a + 2\rho_1 )}  {\nu \rho_2 a}   }  } 
   {2 \sqrt{  -\frac{\rho_1a - \rho_1 - \rho_2 a}{\nu \rho_2 a} }\nu c_4	 + \sqrt{\pi}e^ { -\frac{1}{4} \frac{ \rho_1^2(a-1)^2}{\nu\rho_2 a (\rho_1 a - \rho_1 -\rho_2 a) }  } 
erf \left( \frac{1}{2} \frac{-\eta \rho_2 a + \eta \rho_1 a - \eta\rho_1 - \rho_1 a + \rho_1}{ \nu \rho_2 a \sqrt{-\frac{\rho_1 a -\rho_1 -\rho_2 a}{\nu \rho_2 a}  } }  \right)  
  },
\label{solu_f_eta}
\eqe
where the $erf$ is the usual error function  \cite{NIST}. 

The velocity field of the ideal fluid is the following  
\eq
f = \frac{a-1}{a}g + \frac{c_4}{a}, 
\eqe 
due to the one dimensional property of the model the velocity field of the 
ideal fluid is just scaled by the parameter $a$. 
 
\begin{figure} 
\scalebox{0.4}{
\rotatebox{0}{\includegraphics{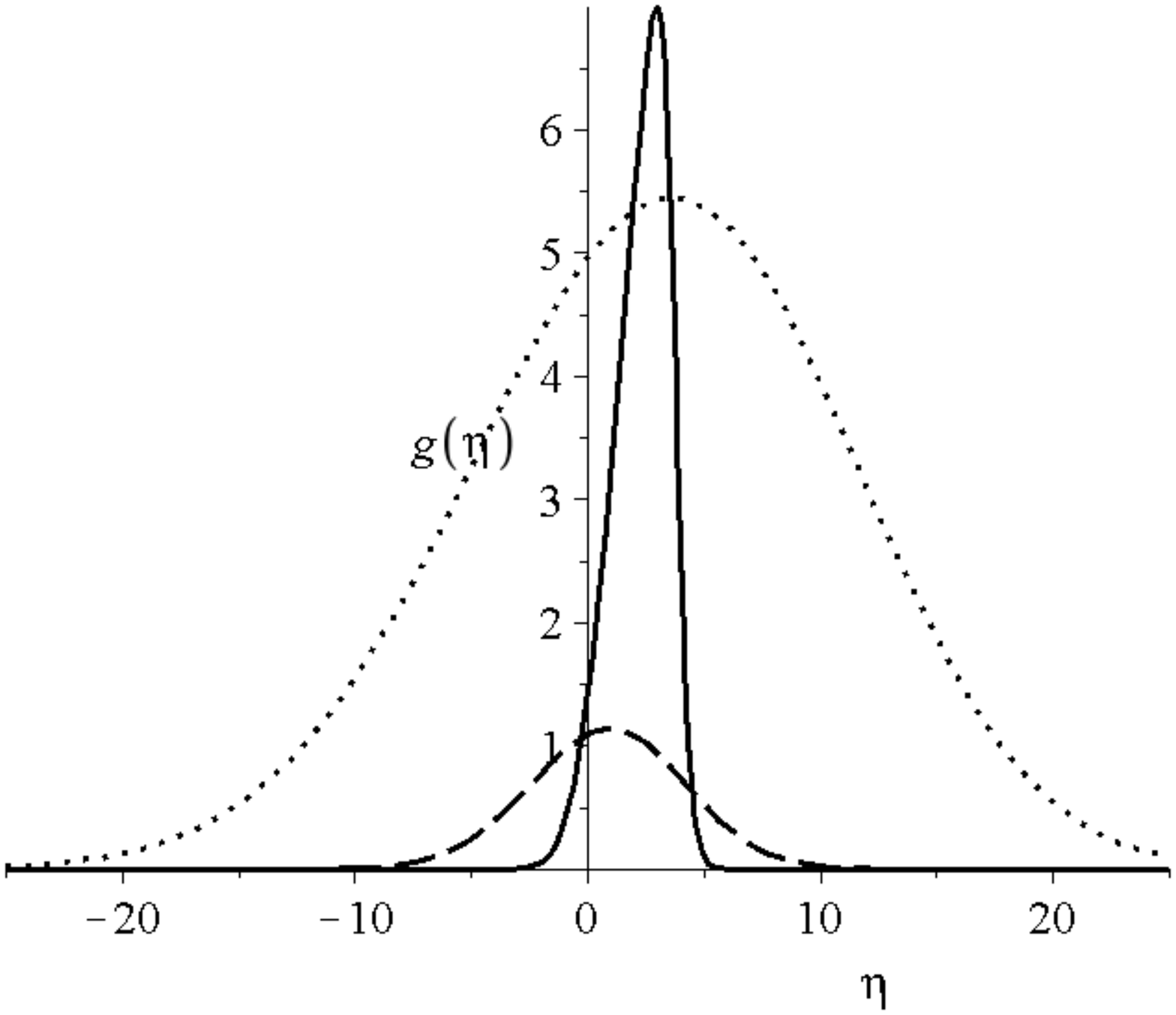}}}
\vspace*{-0.5cm}
\caption{The graphs of Eq.  (\ref{solu_f_eta}) for various parameter sets. The solid curve is for $c_4 = 0.5, a = 0.5, \nu = 1.6, \rho_1 = 2, \rho_2 = 1$ the dotted line is for 
$ c_4 = 0.9, a = 0.27, \nu = 10.4, \rho_1 = 2, \rho_2 = 6$ and the dashed curve is for 
$ c_4 = 0.2, a = 0.27, \nu = 40, \rho_1 = 2, \rho_2 = 20$,  respectively. 
}
\label{egyes}    
\end{figure}

Figure  (\ref{egyes}) shows the shape function of  viscous velocity field for various physical parameters. 
We may say that the functions are zero at large negative arguments, than have a finite value in the origin than have a not-so-pointed maxima and a quick decay to zero at large 
positive arguments.  Different viscosity values, density rates or void fractions cannot modify the general features of the solution.  

It is worth noting here, that the well-known Rayleigh-B\'enard convection model -- which couples the two dimensional Navier-Stokes equation to 
heat conduction -- has a similar solution as well \cite{imre3}. All the velocity, pressure and temperature fields can be expressed with an error function, too. 
Very roughly, we may say that a solution expressed with error functions is almost Gaussian, which means a very sharp temporal and spatial decay. 

For the sake of completeness we have to mention an additional property of the velocity shape function, 
for a small numerical value of $c_4$ the denominator can be zero which means a singularity for the shape function and for the velocity field as well. 
We can see from Eq. (\ref{p}) that the shape function of the pressure field has the same singularity for $c_5 = 0$ as well. 

\begin{figure} 
\scalebox{0.5}{
\rotatebox{0}{\includegraphics{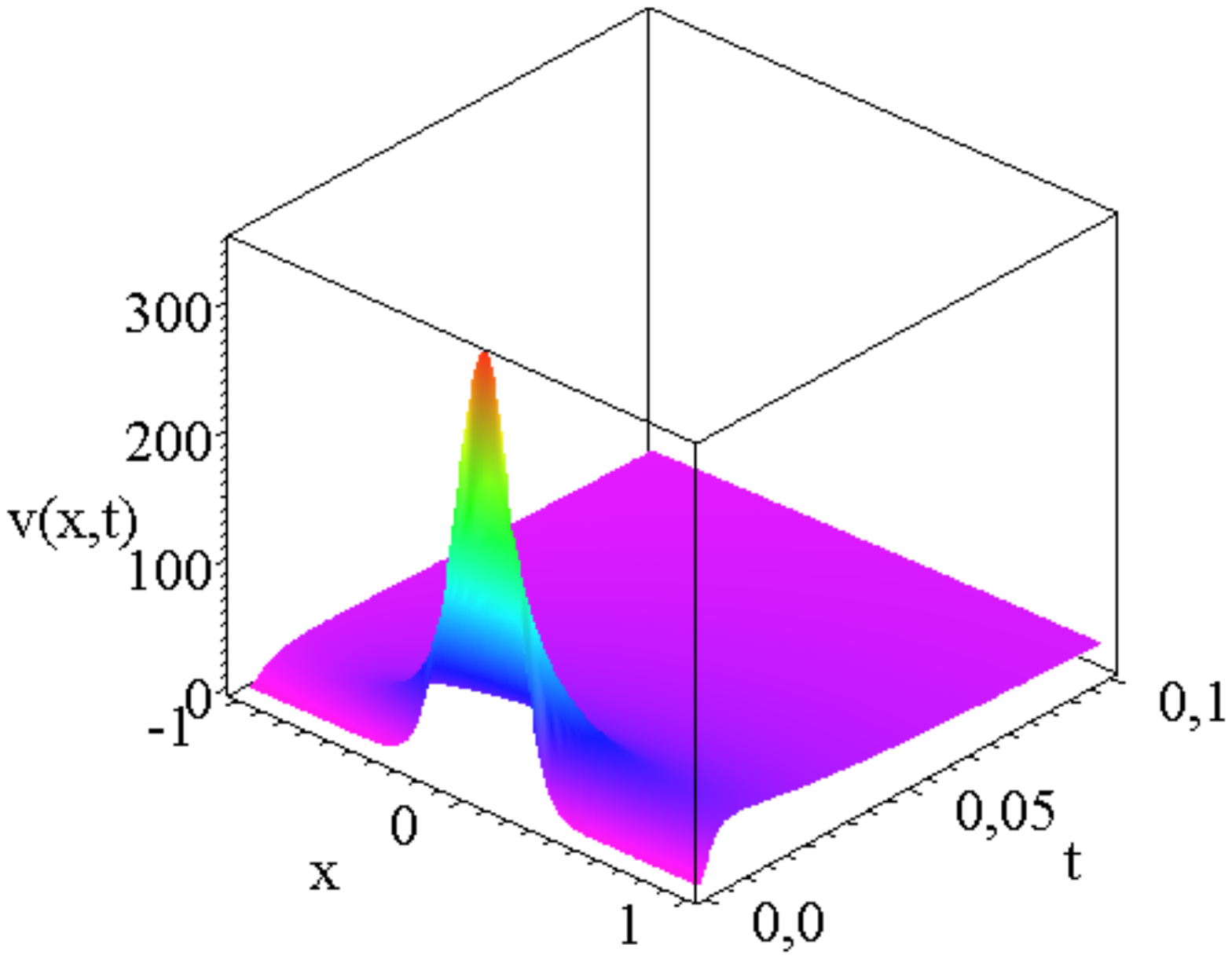}}}
\vspace*{-3.5cm}
\caption{The final velocity distribution of $v_1 = \frac{1}{t^{\frac{1}{2}}}g \left( \frac{x}{t^{\frac{1}{2}}}\right ) $ for the paramter set of 
$ c_4 = 0.2, a = 0.27, \nu = 40, \rho_1 = 2, \rho_2 = 20$.    }
\label{kettes}    
\end{figure}

Figure (\ref{kettes}) presents the final velocity distribution  $v_1(x,t)$ of the  viscous fluid. 
Due to the extra $t^{-\frac{1}{2}}$ prefactor the distribution has an extreme quick time decay.   
It is also interesting, that for a given spatial coordinate, if $t \rightarrow 0$, $\eta \rightarrow 0$, then $g$ has a finite value. This means, that for sufficiently large times at a given $x$, we have 
\begin{equation} 
v_2  \simeq \frac{g(0)}{t^{1/2}}.
\end{equation} 
The situation is similar for the other velocity field. For large times at a given spatial coordinate, a finite $g(0)$ implies a finite $f(0)$. Correspondingly we have 
\begin{equation} 
v_1  \simeq \frac{f(0)}{t^{1/2}}.
\end{equation} 
As one can see, the two velocity field decay in the same rhythm for large times. 


The pressure field is a bit more complicated, but can be evaluated without integration via 
\eq
h= -\frac{\rho_1 f^2}{2} + \frac{\rho_1 \eta f}{2} + c_5\rho_1. 
\label{p}
\eqe
 
\begin{figure} 
\scalebox{0.55}{
\rotatebox{0}{\includegraphics{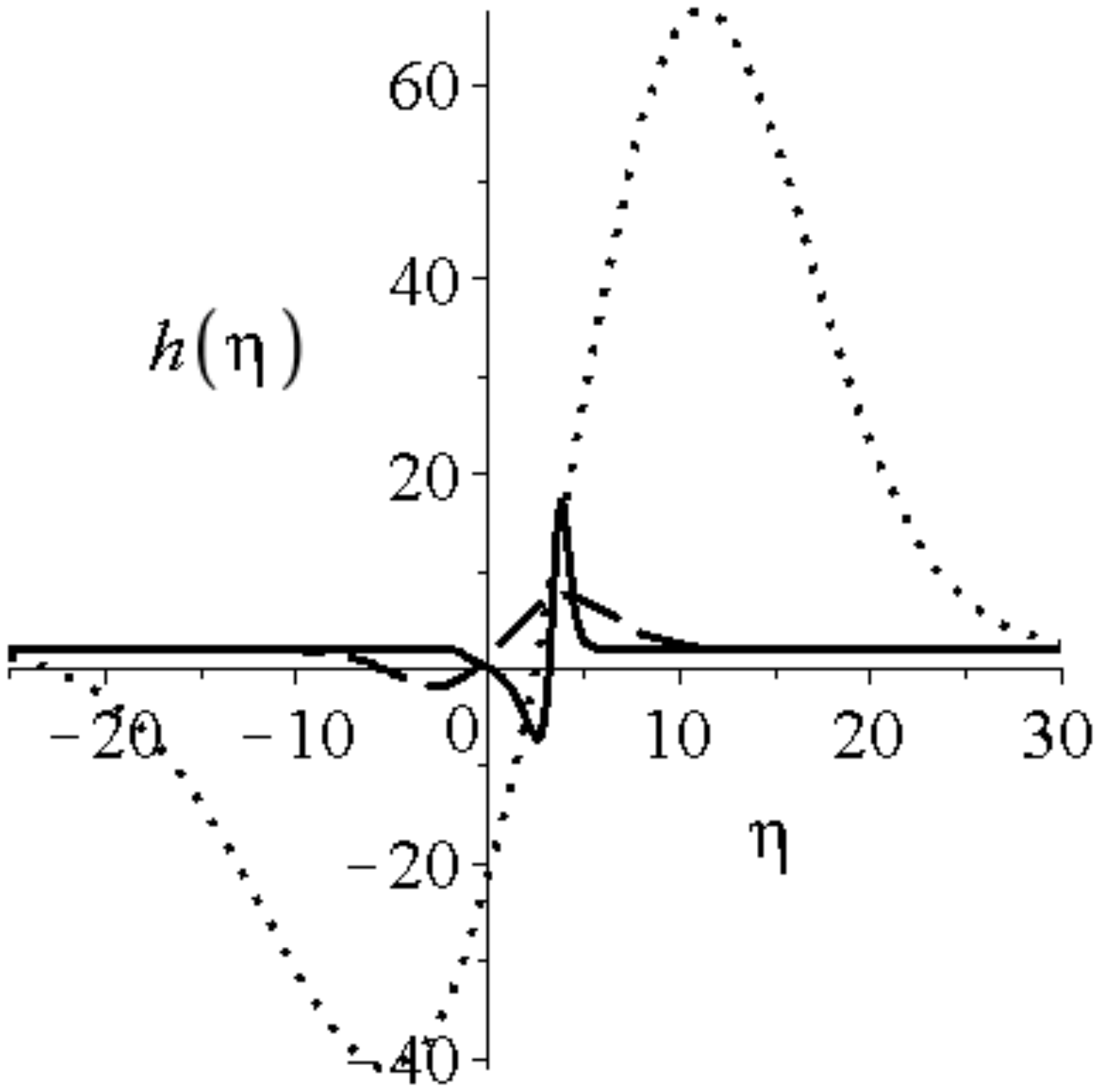}}}
\vspace*{-0.5cm}
\caption{The shape function of the pressure (\ref{p})  for the three parameter sets, given 
above with $c_5 = 0$.}
\label{harmas}        
\end{figure}
\begin{figure} 
\scalebox{0.55}{
\rotatebox{0}{\includegraphics{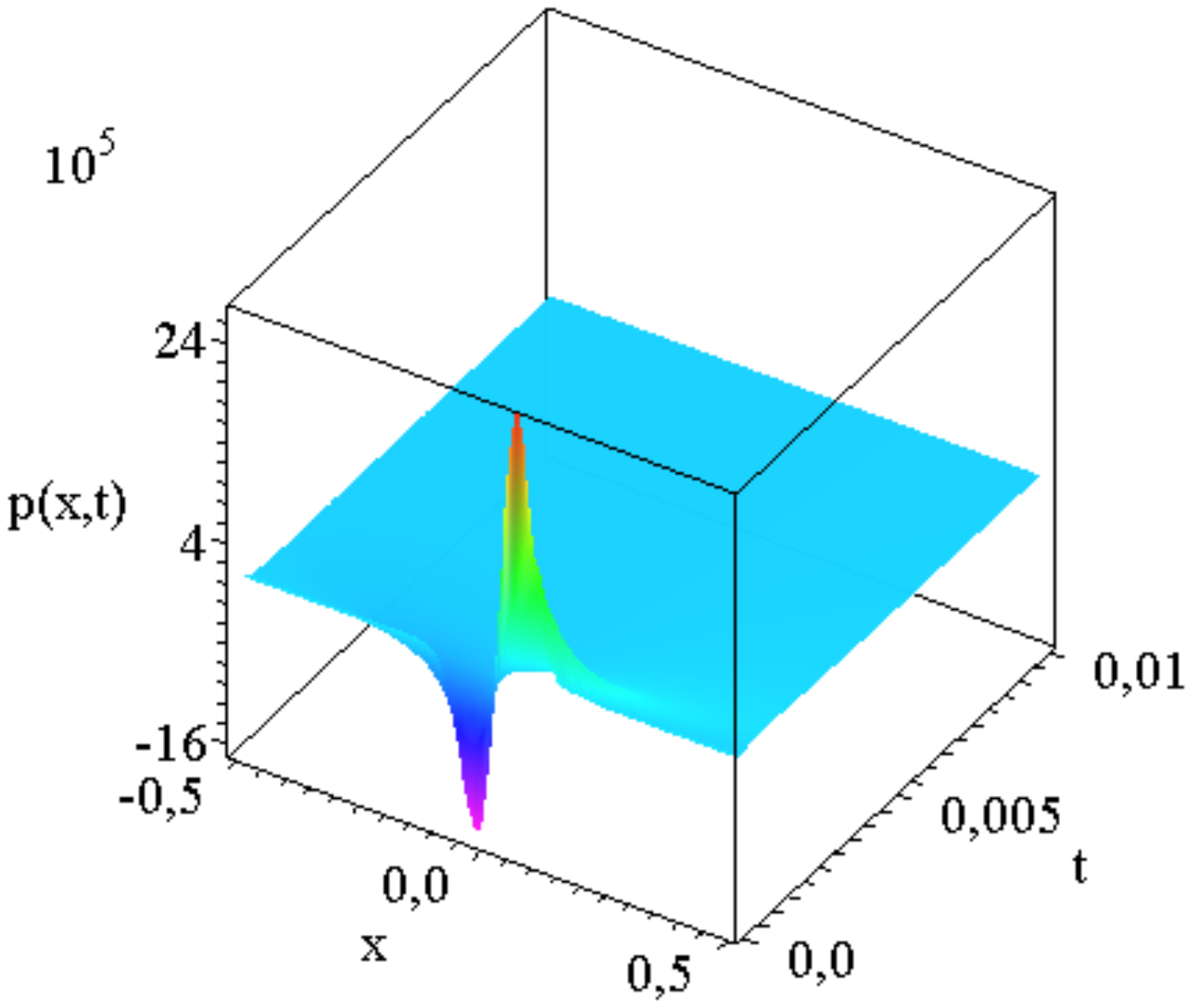}}}
\vspace*{-2.5cm}
\caption{The final pressure distribution of  (\ref{p})   $p = \frac{1}{t^1}h \left( \frac{x}{t^{\frac{1}{2}}}\right ) $   for the parameter set of 
$ c_4 = 0.2, c_5 =0, a = 0.27, \nu = 40, \rho_1 = 2, \rho_2 = 20$.}
We mention that the pressure in eq. (2) is defined up to a constant value. 
If one wants to avoid negative pressures, then an appropriate constant pressure 
can be added to the values presented above, and the shape can be shifted correspondingly. 

\label{negyes}      
\end{figure}
Figure  (\ref{harmas}) shows the pressure shape functions for three different parameter sets. The functions have a bit more complicated structure 
than the velocity distributions, there is a local minima and maxima. 

Our last figure (\ref{negyes}) shows the pressure distribution function. 
Similar to the shape function, it has a very sharp local minima and maxima as well. 
Due to the different exponent $\delta  = 1$ the pressure field has a much quicker 
decay than the velocity field, which is long known for us \cite{imre1}.  

\section{Summary and Outlook} 
After a quick introduction of multi-phase fluids and superconductivity we presented 
the probable most simple one dimensional two-fluid model.
An incompressible Euler and Navier-Stokes equations are coupled together via 
the continuity equations where the divergences of the fluid fields were scaled with the void fraction. 
The self-similar Ansatz was introduced and applied to all dynamical variables of the 
problem. A coupled ODE system was evaluated for the velocity and pressure fields. 
Finally, an analytic solution for the velocity field was derived which contains the error function. 
The pressure field was easily calculated from the velocity field as well.   
Parameter studies were done for both dynamical variables. 
Further work is in progress to enhance the complexity of our present model including 
compressibility with different kind of equation of states or including thermal properties of the corresponding fluids.  
The horizon of the problem can be quite wide.  

\section{Acknowledgment}
This work was supported by project no. 129257 implemented with the
support provided from the National Research, Development and
Innovation Fund of Hungary, financed under the $K \_ 18$ funding
scheme.
 

\begin{references}
 \bibitem{hibiki} M. Ishii and T. Hibiki, {\it{Thermo-Fluid Dynamics of Two-Phase Flow}}, Springer, 2011. 
\bibitem{kolev} N.I. Kolev, {\it{Multiphase Flow Dynamics}}, Springer 2006. 
\bibitem{clayton} C.T. Crowe, {\it{ Multiphase Flow Handbook}}, CRC Taylor and Francis 2006. 
\bibitem{stewart} H.B. Stewart and B. Wendroff, J. Comp.Phys. {\bf{56}}, 363 (1984).
\bibitem{menikoff}R. Menikoff and B. Plohr, Rev.Mod. Phys. {\bf{61}}, 75 (1989).
\bibitem{imre0} I.F. Barna, A.R. Imre, G. Baranyai and Gy. Ezsol, Nuclear Engineering and Design {\bf{240}}, 146 (2010). 
\bibitem{tiselj} I. Tiselj and S. Petelin,  J. Comput. Phys. {\bf{136}}, 503 (1997).
\bibitem{imre_hig} I.F. Barna , A. R. Imre, L. Rosta and F. Mezei    Eur. Phys. J. B {\bf{66}}, 419 (2008). 
\bibitem{schmidt} A. Schmidtt, {\it{Introduction to Superfluidity}}, Springer, 2015. 
\bibitem{khala} I.M. Khalatnikov, {\it{At Introduction to the Theory of Superfluidity}}, Westview Press 2000.   
\bibitem{pit} L. Pitaevskii and S. Stringari, {\it{Bose-Einstein Condensation and Superfluidity}}, Oxford Science Publications, 2015. 
\bibitem{wyatt} R. E. Wyatt, {\it{ Quantum Dynamics with Trajectories: Introduction to Quantum Hydrodynamics}}, Springer, 2005. 
\bibitem{tsubota} M. Tsubota, M. Kobayashi and H. Takeuchi, Phys. Rep. {\bf{522}}, 191 (2013).
\bibitem{madel}  E. Madelung, Naturwissenschaften {\bf{14}}, 1004 (1926).
\bibitem{imre_mad1} I.F. Barna, I.F., Pocsai and L. M\'atyas, 
\textit{Journal of Generalized Lie Theory and Applications},  {\bf{11}}, 1000271 (2017).
\bibitem{imre_mad2}I.F. Barna, M.A. Pocsai and L. M\'aty\'as,  
\textit{Advances in Mathematical Physics},  Article ID 7087295. (2018). 
\bibitem{landau} L. Landau, Physical Review {\bf{60}}, 356 (1941). 
\bibitem{Ba2017} S.\ Balibar,  Laszlo Tisza and the two fluid model of superfluidity,   
C.\ R.\ Physique {\bf 18},  586-591, (2017). 
\bibitem{Ti38} L.\ Tisza,  Nature {\bf 141}, 913 (1938). 
\bibitem{Ti47} L.\ Tisza, Physical Review {\bf 72}(9), 838 (1947) 
\bibitem{mikdal} A. B. Migdal, Nucl. Phys. {\bf{13}}, 655 (1959).
\bibitem{hel} G. E. Volovik, {\it{The Universe in a Helium Droplet}}, Clarendon Press, 2003. 
\bibitem{sedov} L. Sedov, {\it{Similarity and Dimensional Methods in Mechanics}} CRC Press, 1993.
\bibitem{zeldovich} Ya. B. Zel'dovich and Yu. P. Raizer {\it{Physics of Shock 
Waves and High Temperature Hydrodynamic Phenomena}} Academic Press, New York, 1966.
\bibitem{imre1} I.F. Barna, Commun. in Theor. Phys. {\bf{56}}, 745  (2011)
\bibitem{imre2}  I.F. Barna and L. M\'aty\'as,  Fluid. Dyn. Res. {\bf{46,}} 055508 (2014).  
\bibitem{imre3} I.F. Barna and L. M\'aty\'as, Chaos Solitons and Fractals {\bf{78,}}  249 (2015). 
\bibitem{imre4} I.F. Barna, L M\'aty\'as and M.A. Pocsai, Fluid. Dyn. Res. {\bf{52}},  015515 (2020). 
\bibitem{imre_book} D. Campos, {\it{Handbook on Navier-Stokes Equations, Theory and Applied Analysis}}, Nova Publishers, New York, 2017, 
Chapter 16, Page 275 - 304. 
\bibitem{NIST} F.W.J. Olver, D.W. Lozier, R.F. Boisvert and C.W Clark, {\it{NIST handbook of mathematical functions}}, Cambridge University Press, 2010.
\end{references}
\end{document}